\documentclass[twocolumn,showpacs,prb,superscriptaddress]{revtex4}

\usepackage{amsmath,tabularx,graphicx}

\newcommand{\chisg}{\chi_{_\text{SG}}}
\newcommand{\br}{\mathbf{r}}
\newcommand{\bnabla}{\boldsymbol{\nabla}}

\begin{document}

\title{Numerical studies of a one-dimensional three-spin spin-glass model\\
with long-range interactions}

\author{Derek Larson}
\affiliation{Department of Physics, University of California, Santa Cruz,
             California 95064, USA}

\author{Helmut G.~Katzgraber}
\affiliation{Theoretische Physik, ETH Zurich, CH-8093 Zurich,
             Switzerland}
\affiliation{Department of Physics and Astronomy, Texas A\&M University, 
	     College Station, Texas 77843-4242, USA}

\author{M.~A.~Moore}
\affiliation{School of Physics and Astronomy, University of Manchester, 
	    Manchester M13 9PL, United Kingdom}

\author{A.~P.~Young}
\affiliation{Department of Physics, University of California, Santa Cruz, 
             California 95064, USA}

\begin{abstract}

We study a $p$-spin spin-glass model to understand if the
finite-temperature glass transition found in the mean-field regime
of $p$-spin models, and used to model the behavior of structural
glasses, persists in the nonmean-field regime. By using a three-spin
spin-glass model with long-range power-law diluted interactions we
are able to continuously tune the (effective) space dimension via
the exponent of the interactions.  Monte Carlo simulations of the
spin-glass susceptibility and the two-point finite-size correlation
length show that \textit{deep} in the nonmean-field regime, the
finite-temperature transition is lost whereas this is not the case in
the mean-field regime, in agreement with the prediction of Moore and
Drossel [Phys.~Rev.~Lett. {\bf 89}, 217202 (2002)] that three-spin models
are in the same universality class as an Ising spin glass in a magnetic
field. However, \textit{slightly} in the nonmean-field region, we find
an apparent transition in the three-spin model, in contrast to  results
for the Ising spin glass in a field. This may indicate that even larger
sizes are needed to probe the asymptotic behavior in this region.

\end{abstract}


\maketitle

\section{Introduction}
\label{sec:intro}

There has been considerable interest in relating structural glasses
to spin glasses because spin-glass models are more amenable to
analytical and numerical calculations than models of interacting
atoms. This activity was started by Kirkpatrick, Thirumalai, and
Wolynes\cite{kirkpatrick:87b,kirkpatrick:87c,kirkpatrick:87}
who observed a close similarity between the theory for the
dynamics of $p$-spin models with $p > 2$ [at the mean-field (MF)
level] and mode-coupling theory\cite{goetze:92} for the dynamics
of supercooled liquids. At the mean-field level, the $p$-spin model
has two transitions (for a review, see Ref.~\onlinecite{bouchaud:98}).
There is a dynamical transition at a temperature $T = T_d$, also found
in mode-coupling theory, below which ergodicity breaking occurs
but which is not associated with any thermodynamic singularities.
In addition, there is a transition at $T_c < T_d$  which does have
thermodynamic singularities and below which replica symmetry breaking
(RSB) occurs at the ``one-step'' level.\cite{bouchaud:98}
It is this transition which is associated with a possible (ideal)
thermodynamic glass transition of structural glasses, where $T_c$
corresponds to the Kauzmann temperature $T_{\rm K}$.\cite{kauzmann:48}

The connection between structural glasses and $p$-spin models is less
clear beyond the mean-field level. The dynamical transition at $T_d$
is an artifact of the mean-field limit\cite{parisi:99b,bokil:00}
since it arises from an exponentially large number of excited states
which trap the system for exponentially long times, thereby preventing
an infinite system reaching equilibrium. For a finite-dimensional
system, however, activation over \textit{finite} free-energy barriers
restores ergodicity. Thus the only transition which \textit{might}
occur in finite-dimensional $p$-spin models and structural glasses
is the thermodynamic transition at $T_c$.

Even this transition is likely to be significantly different in
finite dimensions from mean-field predictions, especially for odd
$p$. The reason is that odd-$p$ models violate spin-inversion
symmetry ($S_i \to -S_i$ for all $i$; $S_i \in \{\pm1\}$) so
one might expect that the expectation value of the spin would be
nonzero at all temperatures $T$. However, the spin average (and
hence the spin-glass order parameter) is actually zero in mean-field
models because of their infinite connectivity, see, for example,
Ref.~[\onlinecite{castellani:05}].  Nevertheless, in \textit{any}
finite-dimensional models, the spin-glass order parameter would
be nonzero at all $T$ and so any transition must be of the replica
symmetry breaking type. In fact, one of us and Drossel~\cite{moore:02b}
argue that the transition in $p$-spin models with odd $p$ is in
the same universality class as an Ising ($p = 2$) spin glass in a
magnetic field.\cite{almeida:78}

Because models with even $p$ have spin-inversion symmetry, which does
not seem to have an analog in structural glasses, it is natural
to take $p$ odd in order to represent structural glasses.  In the
present paper, we study numerically whether or not a thermodynamic
transition occurs in a $p=3$ spin glass (and hence presumably also
in a structural glass) for a range of space dimensions.

Unfortunately, it is difficult to study spin glasses numerically
in high space dimensions $d$ because the number of spins $N = L^d$
increases rapidly with linear size $L$ and typically one can only study
$N$ of order of a few thousand. Therefore, the range of $L$ is too limited
to perform a finite-size scaling (FSS) analysis.  Recently, it has been
proposed\cite{katzgraber:03,katzgraber:05c,leuzzi:08,leuzzi:08c,katzgraber:09b}
that one can avoid this difficulty by studying a model in one
dimension in which the interactions depend on a power $\sigma$
of the distance.\cite{kotliar:83} Varying $\sigma$ is analogous to
varying $d$ in a finite-dimensional model. In this paper, we consider
values of $\sigma$ corresponding to an effective space dimension
$d_\text{eff}$ both in the mean-field ($d_\text{eff} > 6$) and
nonmean-field ($d_\text{eff} < 6$) regions.  Our main results are
that we find a transition in the mean-field region, and no transition
for $\sigma$ \textit{well} in the nonmean-field region, consistent
with our results\cite{katzgraber:09b,katzgraber:05c} for the Ising
spin glass in a magnetic field. However, for a value of $\sigma$ in
the nonmean-field region, but not far from the critical value below
which mean-field behavior occurs, we find a transition, in contrast
to our results for the Ising case. We shall discuss possible reasons
for this discrepancy.

The paper is structured as follows: in Sec.~\ref{sec:theory}, we give
some theoretical background on the connection between the transition in
the $p > 2$ model and that in the Ising ($p = 2$) model in a magnetic
field. In Sec.~\ref{sec:model}, we define the one-dimensional (1D) three-spin
model and describe the quantities calculated in the simulations. In
Sec.~\ref{sec:numerical}, we briefly give some information on the
numerical method and the parameters of the simulations. Our results
are presented in Sec.~\ref{sec:results} and our conclusions are
summarized in Sec.~\ref{sec:conclusions}.

\section{Theoretical Background}
\label{sec:theory}

The field theory associated with $p$-spin models is a cubic field
theory\cite{kirkpatrick:87b,kirkpatrick:87c,kirkpatrick:87} with
the following Ginzburg-Landau-Wilson Hamiltonian:

\begin{multline}
\mathcal{H}_\text{GLW} = \int d^d \br \, \left\{
{t \over 2} \sum_{\alpha < \beta}  q_{\alpha\beta}^2(\br) + 
{1 \over 2} \sum_{\alpha < \beta}  
\left(\bnabla q_{\alpha\beta}(\br)\right)^2
\right.
\\ 
\left.
- {w_1 \over 6}\, \mathrm{Tr}\, q^3(\br) - {w_2 \over 3}  
\sum_{\alpha < \beta} q_{\alpha\beta}^3(\br) \right\} ,
\label{eq:HGLW}
\end{multline}
where $q_{\alpha\beta}$ is the order parameter and $\alpha$ and $\beta$
are replica indices which run from 1 to $n$, with $n \to 0$.  Terms
of order $q_{\alpha\beta}^4$ and higher have been omitted (and are
``irrelevant'' in the nonmean-field regime). At cubic order there are
two terms and the ratio of their coefficients $R\equiv w_2/w_1$ plays
an important role in the properties of these models at the mean-field
level. When $R > 1$, mean-field theory predicts\cite{gross:85} that
there are two transitions; a dynamical transition at $T_d$ and a
second transition to a state with one-step replica symmetry breaking
at a lower temperature $T_c$.  When $R < 1$, the transitions at $T_d$
and $T_c$ no longer occur; instead there is a single transition to
a state with full RSB (FRSB).

Outside the mean-field limit, one-step replica symmetry breaking,
which occurs in mean field for $R > 1$, is unstable against thermal
fluctuations.\cite{moore:06} As noted in Sec.~\ref{sec:intro},
a FRSB transition, which occurs in mean field for $R < 1$, is in
the same universality class as the Ising spin glass in a magnetic
field.\cite{moore:02b}  Therefore, these arguments imply that the
\textit{only} possible critical point in finite-dimensional $p$-spin
models is in the same universality class as an Ising model in a
magnetic field.

A $p = 3$ model in which the ratio $R$ is less than unity\cite{yeo:PC}
was numerically studied by Parisi, Picco, and Ritort,\cite{parisi:99b}
who found evidence for a transition. When $R<1$, the effective field
in the Ising spin glass in a field mapping is smaller than for $R>1$,
i.e., the correlation length of the system can become very large even
if there is no transition. When the correlation length becomes on the
order of the system size, this finite-size effect can be mistaken for
a genuine phase transition.\cite{moore:02b} It is one of the purposes
of this work to check whether this interpretation of the work of Moore
and Drossel is correct, by studying a $p$-spin long-range model in
one dimension where the interplay between the correlation length and
the system size can be more easily investigated.

\section{Model and Observables}
\label{sec:model}

\begin{figure}[!tbh]
\includegraphics[width=\columnwidth]{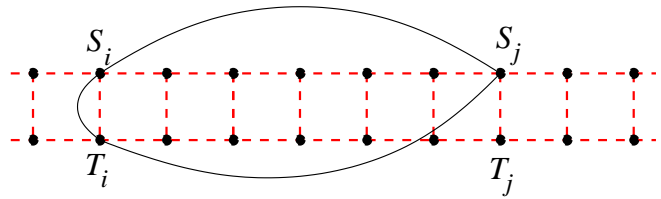}
\caption{(Color online)
One-dimensional three-spin model. The lattice (dashed lines) consists
of a two-leg ladder with an Ising spin $S_i$ at the upper end of the
$i$th rung and an Ising spin $T_i$ at the lower end of the rung. An
interaction couples the two spins at one rung with one of the spins
at another rung. The solid line shows the interaction involving $S_i,
T_i$, and $S_j$.
}
\label{fig:3spinlattice}
\end{figure}

We consider a two-leg ladder with Ising spins $S_i$ and $T_i$ (each
take values $\pm 1$) on each rung, see Fig.~\ref{fig:3spinlattice}.
There are $L$ rungs so $i = 1, \ldots, L$.  Between rungs $i$ and $j$,
one can form four combinations of three spins, namely, $S_i T_i S_j$,
$S_i T_i T_j$, $S_i S_j T_j$, and $T_i S_j T_j$. With a probability
$p_{ij} \sim r_{ij}^{-2\sigma}$, where
\begin{equation}
r_{ij} = (L/\pi)\sin(\pi|i - j|/L)
\end{equation}
is the geometric distance between the spins arranged on a ring, each
of these triplets of spins is coupled by an independent Gaussian random
bond $J_{ij}^{(k)}$ with zero mean and standard deviation unity. With a
$\sigma$-dependent probability $1 - p_{ij}$ they are \textit{all} zero.
To avoid the probability of placing a bond being larger than $1$,
a short-distance cutoff is applied and thus we take
\begin{equation}
p_{ij} = 1 - \exp(-C/r_{ij}^{2\sigma}) \; ,
\label{eq:prob}
\end{equation}
where the constant $C$ is chosen so that the mean coordination number,
\begin{equation}
z = \sum_{j = 2}^{L} p_{1j} 
\end{equation}
takes a fixed value ($z = 6$ here).  The Hamiltonian is therefore given by
\begin{multline} 
{\mathcal H} = -\sum_{i,j} \varepsilon_{ij} \left(
J^{(1)}_{ij} S_i T_i S_j +
J^{(2)}_{ij} S_i T_i T_j + \right. \\
\left.  J^{(3)}_{ij} S_i S_j T_j +
J^{(4)}_{ij} T_i S_j T_j \right)
\; ,
\label{eq:hamiltonian}
\end{multline}
where $\varepsilon_{ij} = 1$ with probability $p_{ij}$ given by
Eq.~\eqref{eq:prob} and zero otherwise.

We now discuss in detail the correspondence between the
long-range one-dimensional model in which $\sigma$ is varied
and a short-range spin-glass model in which the dimension $d$
is varied. This correspondence applies quite generally for
spin-glass models.  By varying $\sigma$, one can tune the model
in Eq.~(\ref{eq:hamiltonian}) from the infinite-range to the
short-range universality classes.\cite{leuzzi:08,katzgraber:09b}
For $0 < \sigma \le 1/2$, the model is infinite range, in the sense
that $\sum_j [J_{ij}^2]_{\rm av}$ diverges, and for $\sigma = 0$, it
corresponds to the Viana-Bray model,\cite{viana:85} i.e., a spin glass
on a random graph.  For $1/2 < \sigma \le 2/3$, the model describes a
mean-field long-range spin glass, corresponding---within the analogy
with short-range systems---to a short-range model with space dimension
above the upper critical dimension $d \ge d_{\rm u} = 6$. For $2/3 <
\sigma \le 1$, the model has nonmean-field critical behavior with
a finite transition temperature $T_c$. For $\sigma \ge 1$, the
transition temperature is zero. We are interested in models which
are not infinite range and which have a finite $T_c$, i.e., $1/2 <
\sigma \le 1$.

A rough correspondence between a value of $\sigma$ in the long-range
1D Ising model and the value of a space dimension
$d_\text{eff}$ in a short-range model can be obtained by comparing
the scaling of the free-energy density, $T_c\, \xi(T,h,d)^{-d}$,
of the $d$-dimensional system to that in the 1D long-range system,
$T_c\, \xi(T,h,\sigma)^{-1}$. When the external field $h$ is zero,
$\xi \sim 1/(T-T_c)^{\nu}$, which gives a matching formula,
\begin{equation}
d_\text{eff}\,\,  \nu_{\rm SR}(d_\text{eff})=\nu_{\rm LR}(\sigma).
\label{eq2:d_sigma}
\end{equation}
A second matching formula\cite{katzgraber:09b} is
\begin{equation}
d_\text{eff} = \frac{2 - \eta_{\rm SR}(d_\text{eff})}{2 \sigma - 1} \, 
\label{eq:d_sigma}
\end{equation}
where $\eta_{\rm SR}(d_\text{eff})$ is the critical exponent $\eta$
for the short-range model, which is zero in the MF regime.
This follows from the dependence of $\xi$ on $h$ at $T_c$, $\xi \sim
h^{-2/(d+2-\eta)}$, and using the fact that for the long-range
system,\cite{kotliar:83}
\begin{equation}
2-\eta_{\rm LR} \equiv 2\sigma-1\quad \text{(MF and non-MF regions)} .
\label{etaLR}
\end{equation}
Equations \eqref{eq2:d_sigma}
and \eqref{eq:d_sigma} agree in the mean-field regime ($d > 6,
1/2 < \sigma < 2/3$), where~\cite{kotliar:83}
\begin{equation}
\nu_{\rm LR} = {1 \over 2\sigma - 1}, \quad \nu_{\rm SR} = \frac{1}{2}\, 
\qquad \text{(MF region)}
\label{nuMFLR}
\end{equation}
and give
\begin{equation}
d_\text{eff} = {2 \over 2 \sigma -1 }\quad \text{(MF region)} \, .
\label{deffMF}
\end{equation}
The aforementioned equations also agree to first order in $6-d$ for
$d<6$ and at the lower critical dimension. Equation (\ref{eq:d_sigma})
has the following required properties: (i) $d_\text{eff} \to \infty$
corresponds to $\sigma \to 1/2$, (ii) the upper critical dimension
$d_{\rm u} = 6$ corresponds to $\sigma_{\rm u} = 2/3$, and (iii) the
lower critical dimension, which is where $d_l - 2 + \eta_{\rm SR}(d_l)
= 0$, corresponds to $\sigma_l = 1$.

To probe the existence of a transition, we compute the
wave-vector-dependent spin-glass susceptibility given by
\begin{widetext}
\begin{equation}
\chisg(k) = \! \frac{1}{L} \sum_{i, j} 
\! \left[\!
\Big( \! \langle S_i S_j\rangle \! - \!
\langle S_i \rangle \langle S_j\rangle \! \Big)^2 
\!\! +
\Big( \! \langle S_i T_j\rangle \!- \! 
\langle S_i \rangle \langle T_j\rangle \! \Big)^2
\!\! + 
\Big( \! \langle T_i S_j\rangle \! - \! 
\langle T_i \rangle \langle S_j\rangle \! \Big)^2
\!\! + 
\Big( \! \langle T_i T_j\rangle \! - \! 
\langle T_i \rangle \langle T_j\rangle \! \Big)^2 
\right]_{\rm av}
\!\!\!\!\!\! 
e^{ik\, (i-j)} ,
\label{eq:chisg}
\end{equation}
\end{widetext}
where $\langle \cdots \rangle$ denotes a thermal average and
$[\cdots]_{\rm av}$ an average over the disorder. To avoid bias,
each thermal average is obtained from a separate copy of the spins.
Therefore, we simulate four copies at each temperature.  Note that
the spin averages $\langle S_i \rangle$ and $\langle T_i \rangle$ are
nonzero even though there is no external field because the interactions
involve three spins and so the model does not have spin-inversion 
symmetry, as discussed in Sec.~\ref{sec:intro}.

The correlation length is given
by\cite{cooper:82,palassini:99b,ballesteros:00,amit:05}
\begin{equation}
\xi_L = \frac{1}{2 \sin (k_\mathrm{m}/2)}
\left[\frac{\chisg(0)}{\chisg(k_\mathrm{m})}
- 1\right]^{1/(2\sigma-1)} ,
\label{eq:xiL}
\end{equation}
where $k_\mathrm{m} = 2 \pi / L$ is the smallest nonzero wave vector
compatible with the boundary conditions.  According to finite-size
scaling,\cite{fss:gtlcd}
\begin{subequations}
\label{eq:xiscale}
\begin{align}
{\xi_L \over L} &= {\mathcal X} [ L^{1/\nu_\text{LR}} (T - T_c) ]  
\;,\;\; (\sigma > 2/3)\, ,
\label{eq:xiscaleNMF}
\\
{\xi_L\over L^{\nu_\text{LR}/3}} &= {\mathcal X} [ L^{1/3} (T - T_c) ] 
\;,\;\; (1/2 <\sigma \le 2/3)\, ,
\label{eq:xiscaleMF}
\end{align}
\end{subequations}
where $\nu_\text{LR}$ is the correlation length exponent, given in the
MF region by Eq.~\eqref{nuMFLR}.  Note, from Eq.~\eqref{eq:d_sigma}
with $\eta_{\rm SR}(d_\text{eff}) = 0$, which is appropriate
for the MF regime, and Eq.~\eqref{nuMFLR}, the power of $L$ in
Eq.~\eqref{eq:xiscaleMF} can be re-expressed in terms of $d_\text{eff}$
according to
\begin{equation}
L^{1/3(2\sigma-1)} \equiv L^{d_\text{eff}/6} ,
\end{equation}
where the factor of 6 occurs because it is the upper critical
dimension $d_u$ for spin glasses. The analogous result for
ferromagnets (for which $d_u = 4$) has been verified numerically
in Ref.~[\onlinecite{jones:05}].  From Eq.~\eqref{eq:xiscale},
if there is a transition at $T = T_c$, data for $ {\xi_L}/{L}$ ($
{\xi_L}/{L^{\nu_\text{LR}/3}}$ in the mean-field region) should cross
at $T_c$ for different system sizes $L$.

We also present data for $\chisg \equiv \chisg(k = 0)$, which
has the finite-size scaling form
\begin{subequations}
\label{eq:chisgscale}
\begin{align}
\chisg &= L^{2 -\eta_\text{LR}} {\mathcal C}[L^{1/\nu_\text{LR}} (T - T_c)]
\;,\;\;\;\; (\sigma > 2/3)\, ,
\label{eq:chisgscaleNMF}
\\
\chisg &= L^{1/3} {\mathcal C}[L^{1/3} (T - T_c)]
\;,\;\;\;\;\; (1/2 <\sigma \le 2/3)\, .
\label{eq:chisgscaleMF}
\end{align}
\end{subequations}
Hence, curves of $\chisg/L^{2 - \eta_\text{LR}}$ ($\chisg /
L^{1/3}$ in the mean-field regime) should also intersect. For
short-range models, Eq.~\eqref{eq:chisgscaleNMF} is less useful
than Eq.~\eqref{eq:xiscaleNMF} in locating $T_c$ because it
involves an unknown exponent $\eta$. However, for long-range
models, $\eta$ is given by Eq.~\eqref{etaLR} {\it exactly} even in
the nonmean-field regime,\cite{fisher:72b,fisher:88,kotliar:83}
and so Eq.~\eqref{eq:chisgscaleNMF} is \textit{just as useful} as
Eq.~\eqref{eq:xiscaleNMF} in this case.\cite{fss:review}

From now on, all exponents will be those of the long-range system
so the subscript $\text{LR}$ will be suppressed.

If there are no corrections to scaling, the intersection
temperatures for all pairs of sizes should be equal to
$T_c$. However, in practice there are corrections to scaling
and the intersection temperatures vary with $L$ and only tend to a
constant for $L \to\infty$. Incorporating the leading correction
to scaling, which is characterized by a universal correction
to scaling exponent $\omega$, the intersection temperature of
data for, e.g., $L$ and $2 L$, $T^\star(L, 2L)$, varies with $L$
as
\begin{equation}
T^\star(L, 2L) = T_c + {A \over L^{\omega + 1/\nu}} \, ,
\label{eq:Tstar}
\end{equation}
where $A$ is a nonuniversal amplitude, see Appendix \ref{appA} and
Refs.~\onlinecite{binder:81b,ballesteros:96a,hasenbusch:08b}.  Equation
\eqref{eq:Tstar} is expected to be valid in the nonmean-field region,
$2/3< \sigma < 1$. Approaching the critical value of $\sigma = 2/3$,
one expects $\omega \to 0$. In the mean-field region, $1/2 < \sigma <
2/3$, the critical exponents are known, but we expect corrections to
Eq.~\ref{eq:Tstar}, as discussed in Appendix~\ref{appA}.

\section{Numerical Method and Equilibration}
\label{sec:numerical}

To speed up equilibration, we use the parallel tempering (exchange)
Monte Carlo method.\cite{hukushima:96} In this approach, one simulates
$N_T$ copies of the spins with the same interactions, each at a
different temperature between a minimum value $T_\text{min}$ and a
maximum value $T_\text{max}$. In addition to the usual single spin flip
moves for each copy, we perform global moves in which we interchange
the temperatures of two copies at neighboring temperatures with a
probability which satisfies the detailed balance condition.  In this
way, the temperature of a particular copy performs a random walk
between $T_\text{min}$ and $T_\text{max}$, thus helping to overcome
the free-energy barriers found in the simulation of glassy systems.

\begin{table}[!tb]
\caption{
Parameters of the simulations for different values of $\sigma$. Here
$N_{\rm samp}$ is the number of samples, $N_{\rm sweep}$ is the
total number of Monte Carlo sweeps, $T_{\rm min}$ and $T_{\rm max}$
are the lowest and highest temperatures simulated, and $N_T$ is the
number of temperatures.  The last column shows the parameter $A$
in Eq.~(\ref{eq:prob}) obtained by fixing $z = 6$ neighbors on average.
\label{tab:simparams}}
\begin{tabular*}{\columnwidth}{@{\extracolsep{\fill}} c r r r r r r l}
\hline
\hline
$\sigma$ & $L$ & $N_\text{samp}$ & $N_\text{sweep}$ & $T_\text{min}$ & $T_\text{max}$ &
$N_{T}$ & $A$
\\
\hline
0.55 &   64 & 10000 &    65536 & 1.25 & 3.40 & 16 & 0.95527 \\
0.55 &  128 & 10500 &   131072 & 1.25 & 3.40 & 16 & 0.81746 \\
0.55 &  256 &  4400 &   524288 & 1.25 & 3.40 & 16 & 0.72314 \\
0.55 &  512 &  3150 &  1048576 & 1.55 & 3.40 & 13 & 0.65411 \\
0.55 & 1024 &   850 &  2097152 & 1.55 & 3.40 & 13 & 0.60129 \\[2mm]

0.75 &   32 &  5000 &    65536 & 0.75 & 3.25 & 11 & 2.02742 \\
0.75 &   64 &  5000 &   131072 & 0.75 & 3.25 & 11 & 1.82345 \\
0.75 &  128 & 16300 &   524288 & 0.75 & 3.25 & 11 & 1.71141 \\
0.75 &  256 &  8500 &  2097152 & 0.75 & 3.25 & 11 & 1.64289 \\
0.75 &  512 &  5600 & 16777216 & 1.00 & 3.50 & 15 & 1.59859 \\
0.75 & 1024 &  1000 & 33554432 & 1.00 & 3.50 & 15 & 1.56903 \\[2mm]

0.85 &   32 &  5000 &    65536 & 0.25 & 4.00 & 23 & 2.65088 \\
0.85 &   64 &  5000 &   262144 & 0.25 & 4.00 & 23 & 2.47900 \\
0.85 &  128 &  4750 &  4194304 & 0.25 & 4.00 & 23 & 2.39485 \\
0.85 &  256 &  3800 & 16777216 & 0.50 & 4.00 & 21 & 2.34867 \\
\hline
\hline
\end{tabular*}
\end{table}

For the simulations to be in equilibrium, the following
equality must hold (see Refs.~[\onlinecite{katzgraber:01}] and
[\onlinecite{katzgraber:09b}]):
\begin{equation}
U = - \frac{4}{T}\,\left[\frac{N_b}{L}\, (1 - \hat{q}_l)
\right]_{\rm av} ,
\label{eq:energyqlh}
\end{equation}
where $U$ is the energy per rung of the ladder, averaged over samples,
\begin{multline}
\hat{q}_l = (4  N_b)^{-1}  \sum_{i < j} \varepsilon_{ij} \left[
\langle S_i S_j \rangle^2 +
\langle S_i T_j \rangle^2 + \right.\\
\left.
\langle T_i S_j \rangle^2 +
\langle T_i T_j \rangle^2 \right]
\end{multline}
is the link overlap of a given sample, and $N_b$ is the number of
pairs of connected sites in that sample (i.e.,~the number of nonzero
values of $ \varepsilon_{ij}$). In the simulations we keep doubling
the number of sweeps until Eq.~\eqref{eq:energyqlh} is satisfied
within error bars.  Note that Eq.~\eqref{eq:energyqlh} refers to an
\textit{average over samples}; the relationship between the energy
and link overlap is \textit{not valid} for individual samples.

\section{Results}
\label{sec:results}

\subsection{$\boldsymbol{\sigma = 0.85}$}
\label{sec:results_0.85}

\begin{figure}[!tbh]
\includegraphics[width=\columnwidth]{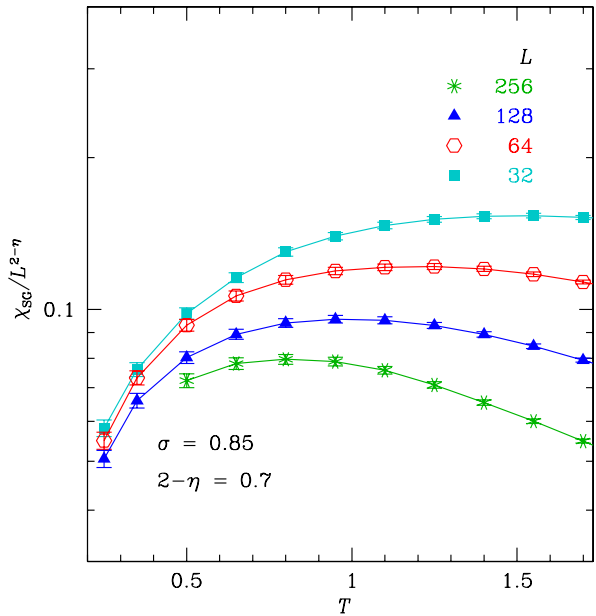}
\caption{(Color online)
Scaled spin-glass susceptibility for $\sigma = 0.85$, in which $2 -
\eta = 2 \sigma - 1 = 0.7$.  According to Eq.~(\ref{eq:chisgscaleNMF}),
the data should intersect at the transition.  The lack of intersections
implies that there is no transition for the studied temperature range.
}
\label{chisg_0.85}
\end{figure}

\begin{figure}[!tbh]
\includegraphics[width=\columnwidth]{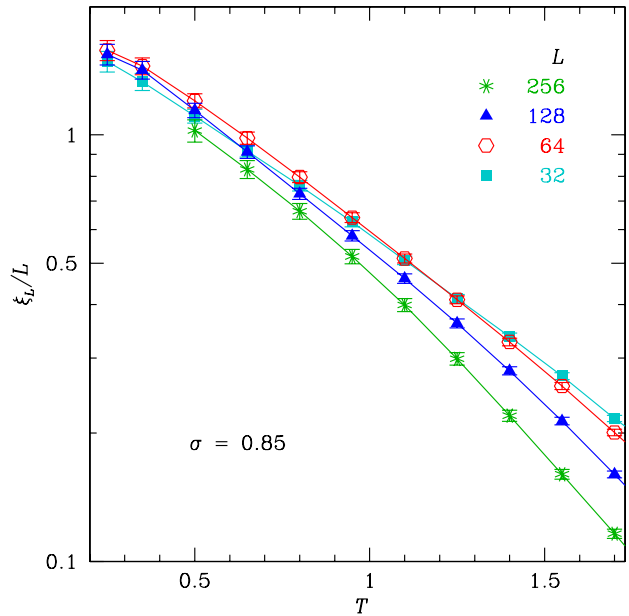}
\caption{(Color online)
Scaled spin-glass correlation length for $\sigma = 0.85$.  According
to Eq.~(\ref{eq:xiscaleNMF}), the data should intersect at the
transition. Although there is an intersection for the smallest pair
of size, there is no intersection for the largest pair, implying
the absence of a transition and in agreement with the data in
Fig.~\ref{chisg_0.85}.
}
\label{xi_L_0.85}
\end{figure}

Results for the spin-glass susceptibility divided by $L^{2-\eta} \equiv
L^{2\sigma - 1} \equiv L^{0.7}$ are shown in Fig.~\ref{chisg_0.85}
for $\sigma = 0.85$ and results for the scaled correlation length
are shown in Fig.~\ref{xi_L_0.85}. The $\chisg$ data show no
intersections (i.e., no sign of a transition). The data for $\xi_L/L$
show an intersection for the smallest pair of sizes, $L= 32$ and
$64$ but no intersection for the largest pair of sizes, $L=128$
and $256$. Hence it appears that for $\sigma = 0.85$, which is well
in the nonmean-field regime, there is no transition. Of course,
we cannot completely exclude a transition at a very low temperature.

\subsection{$\boldsymbol{\sigma = 0.75}$}
\label{sec:results_0.75}

\begin{figure}[!tbh]
\includegraphics[width=\columnwidth]{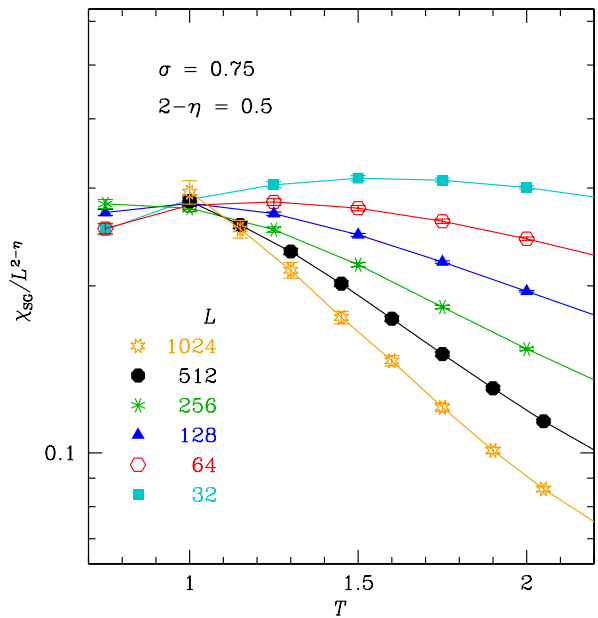}
\caption{(Color online)
Scaled spin-glass susceptibility for $\sigma = 0.75$ in which $2 -
\eta= 2 \sigma - 1 = 0.5$.
}
\label{chisg_0.75}
\end{figure}

\begin{figure}[!tbh]
\includegraphics[width=\columnwidth]{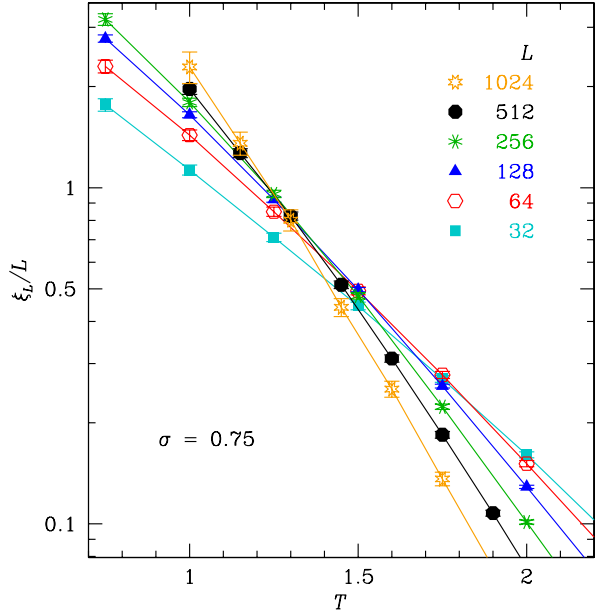}
\caption{(Color online)
Scaled spin-glass correlation length for $\sigma = 0.75$.
}
\label{xi_L_0.75}
\end{figure}

\begin{figure}[!tbh]
\includegraphics[width=\columnwidth]{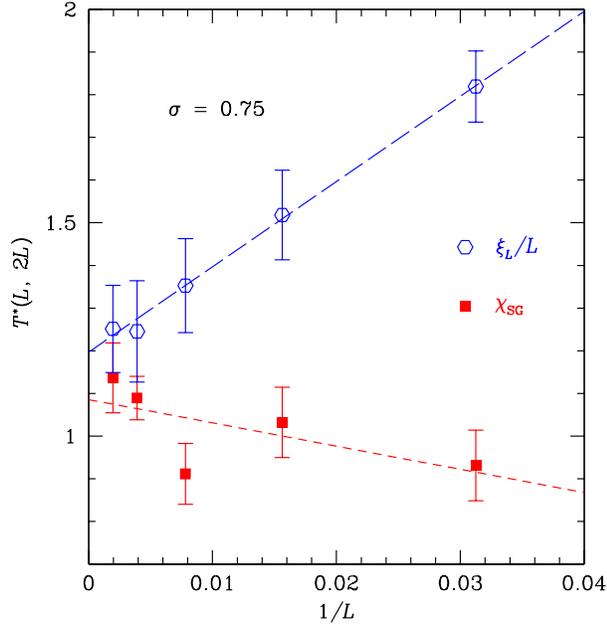}
\caption{(Color online)
Temperatures where data sets for pairs $L$ and $2L$ intersect for
$\sigma = 0.75$.  At large $L$, the data for both $\chisg$ and $\xi_L/L$
extrapolate to a value in the range $1.1$--$1.2$. This implies that
there is a transition at this temperature, unless the true asymptotic
behavior is only seen at even larger sizes.
}
\label{intersec_0.75}
\end{figure}

Our results for $\sigma = 0.75$ are shown in Figs.~\ref{chisg_0.75}
and \ref{xi_L_0.75}. For both $\chisg/L^{2-\eta}$ and $\xi_L/L$,
we find nonzero intersection temperatures $T^\star(L, 2L)$
which are plotted in Fig.~\ref{intersec_0.75}. The horizontal
axis in Fig.~\ref{intersec_0.75} is $1/L$, and, according to
Eq.~\eqref{eq:Tstar}, the data would be a straight line if $1/\nu +
\omega = 1$. Our data are consistent with this but we do not have
good enough data to obtain a precise value for this exponent. The main
point is that, despite strong corrections to scaling, the data for
\textit{both} $\chisg/L^{2-\eta}$ and $\xi_L/L$ indicate a transition
with $T_c$ in the range from $1.1$ to $1.2$.

This is rather surprising since it has been argued\cite{moore:02b}
that the transition is in the same universality class as the Ising
spin glass in a magnetic field, and \textit{no} transition has
been found for that model with $\sigma = 0.75$ in work by some of
us.\cite{katzgraber:09b,katzgraber:05c} However, corrections to scaling
are very large (see Figs.~\ref{chisg_0.85}--\ref{intersec_0.55}),
and so it is plausible that system sizes considerably larger
than $L = 1024$ are needed to see the true thermodynamic
behavior of the three-spin model when $\sigma \searrow 2/3$, in
which case there would be no inconsistency with the work of
Refs.~[\onlinecite{katzgraber:09b}] and [\onlinecite{katzgraber:05c}].

\subsection{$\boldsymbol{\sigma = 0.55}$}
\label{sec:results_0.55}

\begin{figure}[!tbh]
\includegraphics[width=\columnwidth]{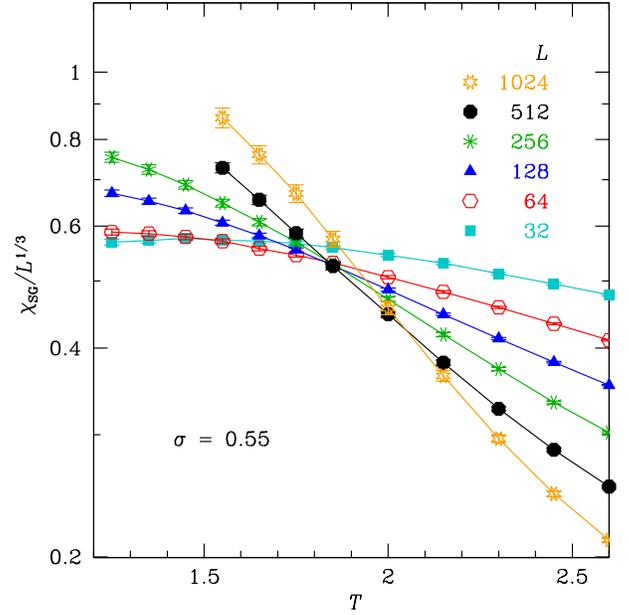}
\caption{(Color online)
Scaled spin-glass susceptibility for $\sigma = 0.55$ according to
Eq.~(\ref{eq:chisgscaleMF}).  The data are consistent with a transition
at $T_c \simeq 2.1$, see also Fig.~\ref{intersec_0.55}.
}
\label{chisg_0.55}
\end{figure}

\begin{figure}[!tbh]
\includegraphics[width=\columnwidth]{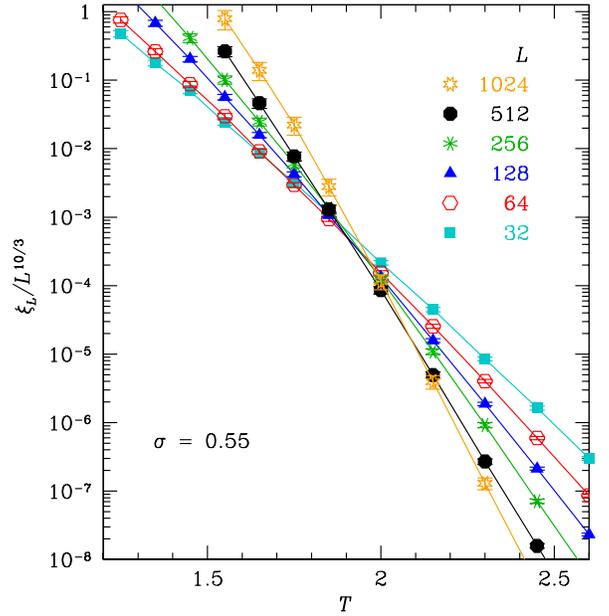}
\caption{(Color online)
Scaled correlation length for $\sigma = 0.55$.  The data are
consistent with a transition at $T_c \simeq 2.1$, see also
Fig.~\ref{intersec_0.55}.
}
\label{xi_L_0.55}
\end{figure}

\begin{figure}[!tbh]
\includegraphics[width=\columnwidth]{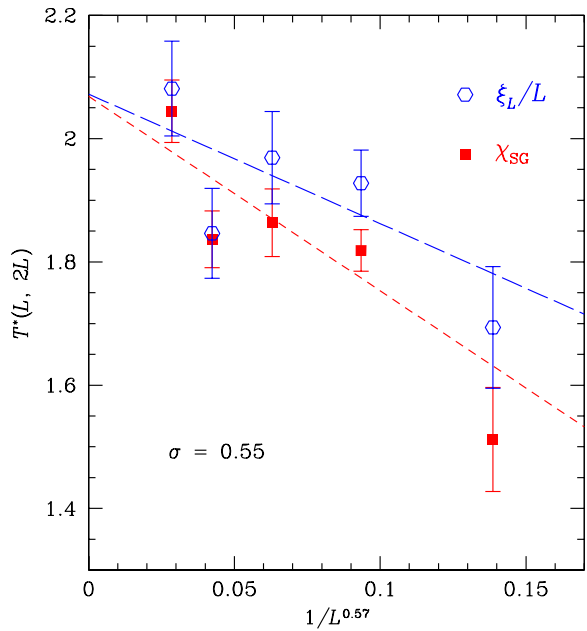}
\caption{(Color online)
Temperatures where data for pairs $L$ and $2L$ intersect for $\sigma
= 0.55$.  At large $L$, the data for both $\chisg$ and $\xi_L/L$
extrapolate to a value of approximately $2.1$, implying that there
is a transition at this temperature.
}
\label{intersec_0.55}
\end{figure}

Our results for $\sigma = 0.55$ (mean-field regime) are shown in
Figs.~\ref{chisg_0.55} and \ref{xi_L_0.55}. As discussed in Appendix
\ref{appA}, the intersection temperatures in the mean-field regime
are given by Eq.~\eqref{Tstar2}. For $\sigma = 0.55$, the exponent
$5/3 - 2\sigma$ is equal to 0.57. We therefore plot the intersection
temperatures against $1/L^{0.57}$ in Fig.~\ref{intersec_0.55}.
The data strongly suggest that there is a transition at $T_c \simeq
2.1$.  This result is consistent with our earlier results for the Ising
spin glass in a magnetic field,\cite{katzgraber:09b,katzgraber:05c}
where we also found a transition in the mean-field region.

\section{Summary and Conclusion}
\label{sec:conclusions}

We have studied the existence of phase transitions in a three-spin
spin-glass model, that is argued to be an appropriate model to describe
the (possible) ideal glass transition in a supercooled liquid. We
have studied three values of the parameter $\sigma$: (i) $\sigma
= 0.55$ (mean-field regime), (ii) $\sigma = 0.75$ (nonmean-field
region, but close to the mean-field boundary at $2/3$), and (iii)
$\sigma = 0.85$ (deep inside the nonmean-field regime). Moore and
Drossel~\cite{moore:02b} argue that any transition in this model is
in the same universality class as that of the Ising spin glass in a
magnetic field. In particular, the two models should have the same
critical value of sigma where the transition disappears (corresponding
to the lower critical dimension for the short-range case). In other
words, if one model has a transition the other should have one
and vice versa.

For the mean-field case, $\sigma = 0.55$, we find a
finite-temperature transition.  Comparing with our previous
work\cite{katzgraber:09b,katzgraber:05c} for the Ising spin glass in a
magnetic field, in which we also find a transition in the mean-field
regime, this result is seen to be consistent with the predictions
of Ref.~\onlinecite{moore:02b}.

For the case studied that is \textit{well} in the nonmean-field
regime, $\sigma = 0.85$, we find no transition, in agreement with
our work for the Ising spin glass in a magnetic field. This implies
that there is no ideal glass transition in three dimensions since
$d=3$ is \textit{well} below the upper critical dimension of $d=6$ for
models with cubic interactions, such as that in Eq.~(\ref{eq:HGLW}).

However, for $\sigma = 0.75$ the results presented here,
which indicate a finite transition temperature, appear to be
at odds with our results for the Ising transition in a field,
\cite{katzgraber:09b,katzgraber:05c} where we find no transition. We
note, however, that Leuzzi {\em et al}.~\cite{leuzzi:08c} argue that
there \textit{is} a transition for this case, based on a nonstandard
finite-size scaling analysis.  In the absence of a transition, the
system breaks up into domains of size $\ell$ (Imry-Ma length) which
can be large at low temperatures, depending on the model. A possible
explanation of our results for $\sigma = 0.75$ is that $\ell(T \to 0)$
is greater than the largest system size, namely, $L = 1024$, for the
three-spin model, although not for the Ising model in a field studied in
Ref.~[\onlinecite{katzgraber:09b}]. If this is the case, even larger
values of $L$ are needed to determine the asymptotic behavior of the
three-spin model.

\begin{acknowledgments}

Most of the simulations were performed on the ETH Z\"urich brutus
cluster. H.G.K.~acknowledges support from the Swiss National Science
Foundation under Grant No.~PP002-114713 and would like to thank the
ETH Zurich Center for Theoretical Studies for financial support of
D.L.~during a visit to ETH Zurich.  A.P.Y.~acknowledges support from
the NSF under Grant No.~DMR-0906366. We are also grateful for a generous
allocation of computer time from the Hierarchical Systems Research
Foundation.

\appendix
\section{Size dependence of Intersection Temperatures}
\label{appA}

According to standard finite-size scaling, the spin-glass
susceptibility varies near the critical point according to
\begin{equation}
\chisg(t, L) = L^a\left [ f(L^b t) + L^{-\omega} g(L^y t) + 
\cdots \right] + c_0 + c_1 t + \cdots ,
\label{fss}
\end{equation}
where $t = T - T_c$.  The $L^{-\omega}$ term is the leading
\textit{singular} correction to scaling and $c_0$ is the leading
\textit{analytic} correction to scaling.

\textbf{Nonmean-field regime}. In the nonmean-field regime,
$\sigma_c < \sigma < 1$ with $\sigma _c = 2/3$, we have $a = 2 -\eta
= 2 \sigma - 1$ and $b = 1 / \nu $. (In this section all exponents
refer to the long-range universality class.) We use Eq.~\eqref{fss}
to calculate the temperature $T^\star(L, 2L)$ where data for $\chisg /
L^a$ for sizes $L$ and $2L$ intersect. Expanding $f(x)$ to first order
in $x$, replacing $g(x)$ by $g(0)$, and assuming that $a > \omega$
(which is certainly true near $\sigma = 2/3$, where $\omega \to 0$)
we recover Eq.~\eqref{eq:Tstar}.

\textbf{Mean-field regime}.  Curiously, the situation in
the mean-field regime, $1/2 < \sigma \le 2/3$, is more
complicated. First of all, the exponents $a$ and $b$ are
independent of $\sigma$\cite{binder:85,luijten:99,jones:05}
and take the value at $\sigma_c$ for all $1/2 < \sigma <
\sigma_c$, i.e. $a=b= 1/3$.  Second, although the $L^{2\sigma
-1}$ term is replaced as the \textit{largest} term by an
$L^{1/3}$ term (due to the presence of a ``dangerous irrelevant
variable,''cf.~Refs.~[\onlinecite{binder:85,luijten:99,jones:05}])
we expect this term to not disappear but rather become a
\textit{correction to scaling}. Hence, we replace Eq.~\eqref{fss} by
\begin{multline}
\chisg(t, L) = L^{1/3}\left [ f(L^{1/3} t) + L^{-\omega} g(L^{1/3} t) +
\cdots \right] \\
+ d_0 L^{2\sigma-1} hg(L^{1/3} t) + c_0 + \cdots \quad (1/2 < \sigma < 2/3) .
\label{fss2}
\end{multline}
The correction exponent $\omega$ can be obtained in the mean-field
regime from the work of Kotliar {\em et al}.~\cite{kotliar:83} and
is~\cite{omega} given by $\omega = 2 - 3 \sigma$.

For $\sigma < \sigma_c$, we find that the $L^{2\sigma -1}$ term
gives the leading correction in Eq.~\eqref{fss2} and, as a result,
Eq.~\eqref{eq:Tstar} is replaced by
\begin{equation}
T^\star(L, 2L) = T_c + {A' \over L^{5/3\, -\, 2\sigma}},
\qquad (1/2 < \sigma < 2/3)\, .
\label{Tstar2}
\end{equation}

To determine the intersection temperatures of the correlation length
we also need the FSS scaling form for $\chisg(k_\mathrm{m})$. We find
that there is an additional correction which dominates for $\sigma <
7/12 = 0.5833$. However, for the value $\sigma = 0.55$ used in the
simulations, the resulting difference from Eq.~\eqref{Tstar2} is very
small and therefore we neglect it.

\end{acknowledgments}

\bibliography{refs,comments}

\begin{thebibliography}{41}
\bibitem[{fss({\natexlab{a}})}]{fss:gtlcd}
\bibinfo{note}{For a discussion of how finite-size scaling is modified in the
  region of mean-field exponents, see for example
  Refs.~[\onlinecite{brezin:82,brezin:85,binder:85,luijten:99,jones:05}].}
\bibitem[{ome()}]{omega}
\bibinfo{note}{This expression vanishes for $\sigma = \sigma_c = 2/3$ as
  expected. Furthermore, using the connection between $\sigma$ and the
  effective dimension of a short-range model in Eq.~\eqref{deffMF}, we have
  $\omega = \omega_{\rm SR}/d$ with $\omega_{\rm SR} = (d - 6)/2$, as expected
  at the trivial Gaussian fixed point in a cubic field
  theory.\cite{harris:76}}.
\end{thebibliography}

\end{document}